\documentclass[runningheads]{llncs}
\usepackage[T1]{fontenc}
\usepackage{graphicx}
\usepackage{subfigure}
\usepackage{array}
\usepackage{diagbox}
\usepackage[graphicx]{realboxes}
\usepackage{cite}
\usepackage{float}
\usepackage{marvosym}
 \usepackage{mathrsfs}
 \usepackage{amsmath}
\usepackage{amssymb}
\usepackage{amsmath}
\usepackage[colorlinks=true,
            linkcolor=blue,
            anchorcolor=blue,
            citecolor=blue,
            urlcolor=blue,
            ]{hyperref}
\begin{document}
\title{UWAT-GAN: Fundus Fluorescein Angiography Synthesis via Ultra-wide-angle Transformation Multi-scale GAN}
\titlerunning{UWAT-GAN for Fundus Fluorescein Angiography Synthesis}
%
\author{Zhaojie Fang\inst{1} \and
Zhanghao Chen\inst{1} \and
Pengxue Wei\inst{3} \and
Wangting Li\inst{3} \and
Shaochong Zhang\inst{3} \and
Ahmed Elazab\inst{4} \and
Gangyong Jia\inst{2} \and
Ruiquan Ge\inst{2}\textsuperscript{\Letter} \and
Changmiao Wang\inst{5}\textsuperscript{\Letter}}

%
%
\authorrunning{Z. Fang et al.}
%
\institute{HDU-ITMO Joint Institute, Hangzhou Dianzi University, Hangzhou 310018, China \and
School of Computer Science and Technology, Hangzhou Dianzi University, Hangzhou 310018, China \\
\email{gespring@hdu.edu.cn}
\and
Shenzhen Eye Hospital, Jinan University, Shenzhen 518040, China \and
School of Biomedical Engineering, Shenzhen University, Shenzhen 518060, China \and
Medical Big Data Lab, Shenzhen Research Institure of Big Data, Shenzhen 518172, China \\
\email{cmwangalbert@gmail.com}}

\maketitle              
\begin{abstract}
Fundus photography is an essential examination for clinical and differential diagnosis of fundus diseases. Recently, Ultra-Wide-angle Fundus (UWF) techniques, UWF Fluorescein Angiography (UWF-FA) and UWF Scanning Laser Ophthalmoscopy (UWF-SLO) have been gradually put into use. However, Fluorescein Angiography (FA) and UWF-FA require injecting sodium fluorescein which may have detrimental influences. To avoid negative impacts, cross-modality medical image generation algorithms have been proposed. Nevertheless, current methods in fundus imaging could not produce high-resolution images and are unable to capture tiny vascular lesion areas. This paper proposes a novel conditional generative adversarial network (UWAT-GAN) to synthesize UWF-FA from UWF-SLO. Using multi-scale generators and a fusion module patch to better extract global and local information, our model can generate high-resolution images. Moreover, an attention transmit module is proposed to help the decoder learn effectively. Besides, a supervised approach is used to train the network using multiple new weighted losses on different scales of data. Experiments on an in-house UWF image dataset demonstrate the superiority of the UWAT-GAN over the state-of-the-art methods. The source code is available at: https://github.com/Tinysqua/UWAT-GAN.

\keywords{Fluorescein Angiography \and Cross-modality Image Generation \and Ultra-Wide-angle Fundus Imaging \and Conditional Generative Adversarial Network.}
\end{abstract}
\section{Introduction}
Fluorescein Angiography (FA) is a commonly utilized imaging modality for detecting and diagnosing fundus diseases. It is widely used to image vascular structures and dynamically observe the circulation and leakage of contrast agents in blood vessels. Recently, the emergence of Ultra-Wide-angle Fundus (UWF) imaging has enabled its combination with FA and Scanning Laser Ophthalmoscopy (SLO), namely UWF-FA and UWF-SLO. The UWF-FA imaging enables simultaneous and high-contrast angiographic images of all 360 degrees of the mid and peripheral retina \cite{ref_1,ref_4,ref_21}. However, both FA and UWF-FA require injecting a fluorescent dye (i.e., sodium fluorescein) into the anterior vein of the patient's hand or elbow, which then passes through the blood circulation to the fundus blood vessels. Some patients may experience adverse reactions such as vomiting and nausea during or after the examination. Moreover, it is not suitable for patients with serious cardiovascular and other systemic diseases.\par
Cross-modality medical image generation provides a new method for solving the aforementioned problems. Multi-scale feature maps from different input modalities usually have similar structures. Hence, different contrasts can be merged to generate target images based on multimodal deep learning to provide more information for diagnosis \cite{ref_15}. Recently, the generative adversarial networks (GANs) \cite{ref_5} and their variants have made breakthroughs in this field. The idea of PatchGAN \cite{ref_12} was proposed to synthesize clearer images. Liu et al. \cite{ref_14} proposed an end-to-end multi-input and multi-output deep adversarial learning network for MR image synthesis. Xiao et al. \cite{ref_22} proposed a Transfer-GAN model by combining transfer learning and GANs to generate CT high-resolution images. By merging multi-scale generators, these networks can explore fine and coarse features from images \cite{ref_10}. Kamran et al. \cite{ref_9} proposed a semi-supervised model called VTGAN introducing transformer module into discriminators, helping the synthesis of vivid images. However, previous methods yielded lower-resolution situations and most discriminators can only take squared inputs (width equal to height)\cite{ref_18}. Moreover, misaligned data and lower attention on disease-related regions as well as the correctness of synthesized lesions remain significant issues. Furthermore, the highly non-linear relationship between different modalities makes the mapping from one modality to another difficult to learn \cite{ref_23}.\par
In this paper, we present the Ultra-Wide-Angle Transformation GAN (UWAT-GAN), a supervised conditional GAN capable of generating UWF-FA from UWF-SLO. To address the image misalignment issue, we employ an automated image registration method and integrate the idea of pix2pixHD \cite{ref_20} to use multi-scale discriminators. In addition, we use the multi-scale generators to synthesize high-resolution images as well as improve the ability to capture tiny vascular lesion areas and employ multiple new weighted losses on different scales of data to optimize model training. For evaluation metrics, we use Fréchet Inception Distance (FID) \cite{ref_6}, Kernel Inception Distance (KID) \cite{ref_11}, Inception Score (IS) \cite{ref_2} and Learned Perceptual Image Patch Similarity (LPIPS) \cite{ref_24} to quantify the quality of images. Finally, we compare  UWAT-GAN with the state-of-the-art image synthesis frameworks \cite{ref_7, ref_20, ref_3} for qualitative assessment and conduct an ablation study. Our main contributions are summarised as follows:\par
1). To the best of our knowledge, we present the first study to synthesize UWF-FA from UWF-SLO, overcoming the limitations of UWF-FA imaging.\par
\begin{figure}[t]
\centering 
\includegraphics[height=5.6cm,width=12.0cm]{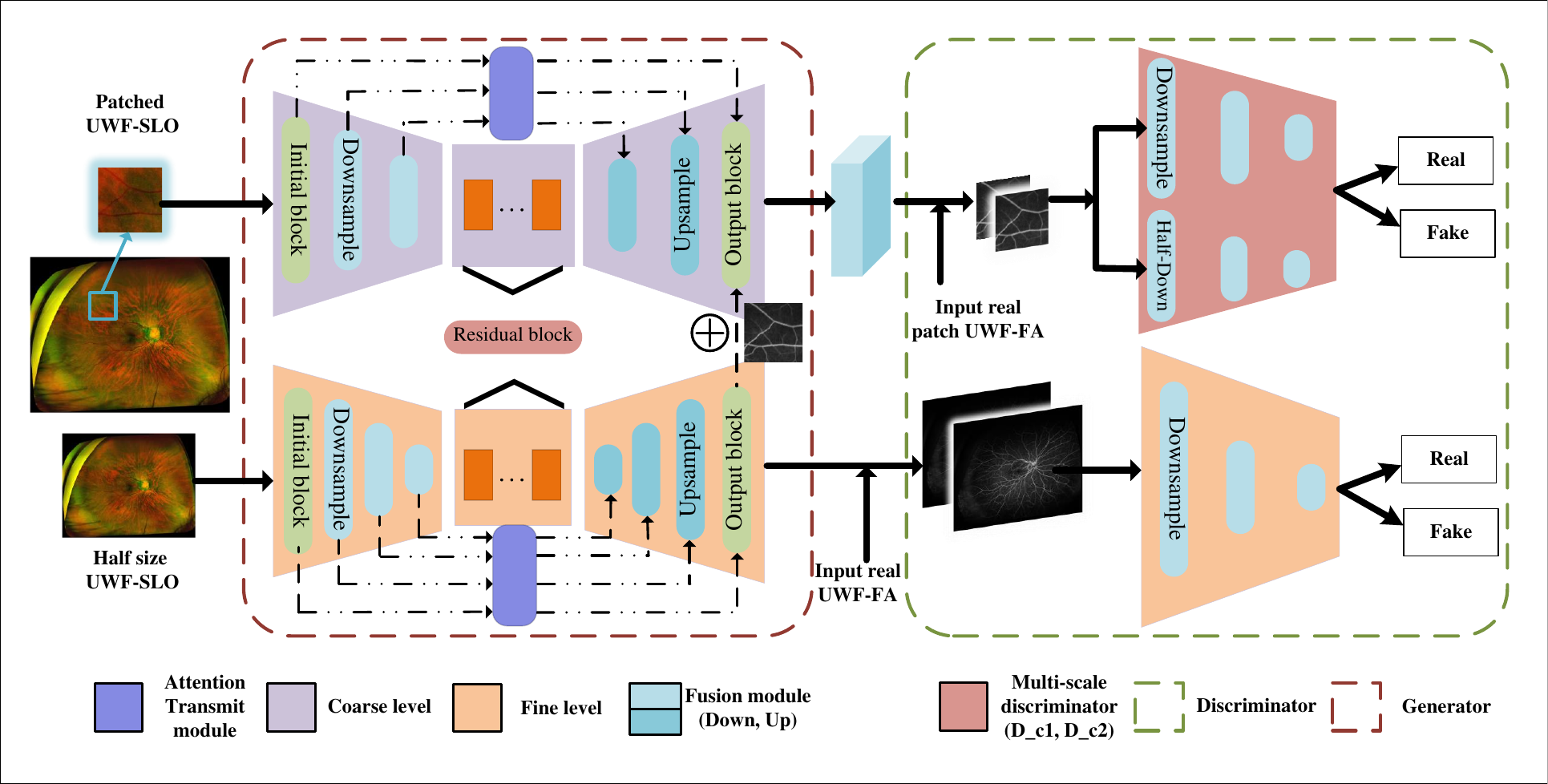}
\caption{The overall framework of UWAT-GAN.}
\label{1}
\end{figure}
2). We propose a novel UWAT-GAN utilizing multi-scale generators and multiple new weighted losses on different data scales to synthesize high-resolution images with the ability to capture tiny vascular lesion areas. \par
3). We assess the performance of the UWAT-GAN on a clinical in-house dataset and adopt an effective preprocessing method for image sharpening and registration to enhance the clarity of vascular regions and tackle the misalignment problem.\par
4). We demonstrate the superiority of the proposed UWAT-GAN against the state-of-the-art models through extensive experiments, comparisons, and ablation studies.

\section{Methodology}
We propose a supervised conditional GAN for synthesizing UWF-FA from UWF-SLO images, as illustrated in Fig. \ref{1}. In order to achieve the desired outcome, we propose the concept of a fine-coarse level generator in whole architecture (Sec. \ref{S1}) and a fusion module works on result of both level generators (Sec. \ref{S2}). Then the Attention Transmit Module is put forward to improve the U-Net-like architecture (Sec. \ref{S3}). Additionally, we provide a comprehensive description of up-down sampling process and architecture of multi-scaled discriminators (Sec. \ref{S4}). Eventually, we discuss the proposed loss function terms and their impacts in detail (Sec. \ref{S5}).
\subsection{Overall Architecture} \label{S1}
UWF-FA has global features such as eyeballs and long-thick blood vessels, as well as local features such as small lesions and capillary blood vessels. However, generating images with both global and local features using a single generator is challenging. To address this issue, we devise two different levels of generators. The coarse generator $Gen_C$ extracts global information and generates a result based on this information, while the fine generator $Gen_F$ extracts local information. The results of global and local information can be used, alternately, as a reference for each other. Hence, this allows the extraction and utilization of both global and local information. In Fig. \ref{1}, the original image is fed into the entire model. After down-sampling, the image is passed into $Gen_C$. Then, we extract a patch from the original image as an input to $Gen_F$ as described in Sec. \ref{S3}. Both generators share the down-sample residual block and attention concatenated modules. Note that both generators generate a UWF-FA image and pass it to the discriminators. However, the output of $Gen_F$ is the one we considered in the later experiments.

\begin{figure}[t]
\centering 
\includegraphics[height=4.32cm,width=12.32cm]{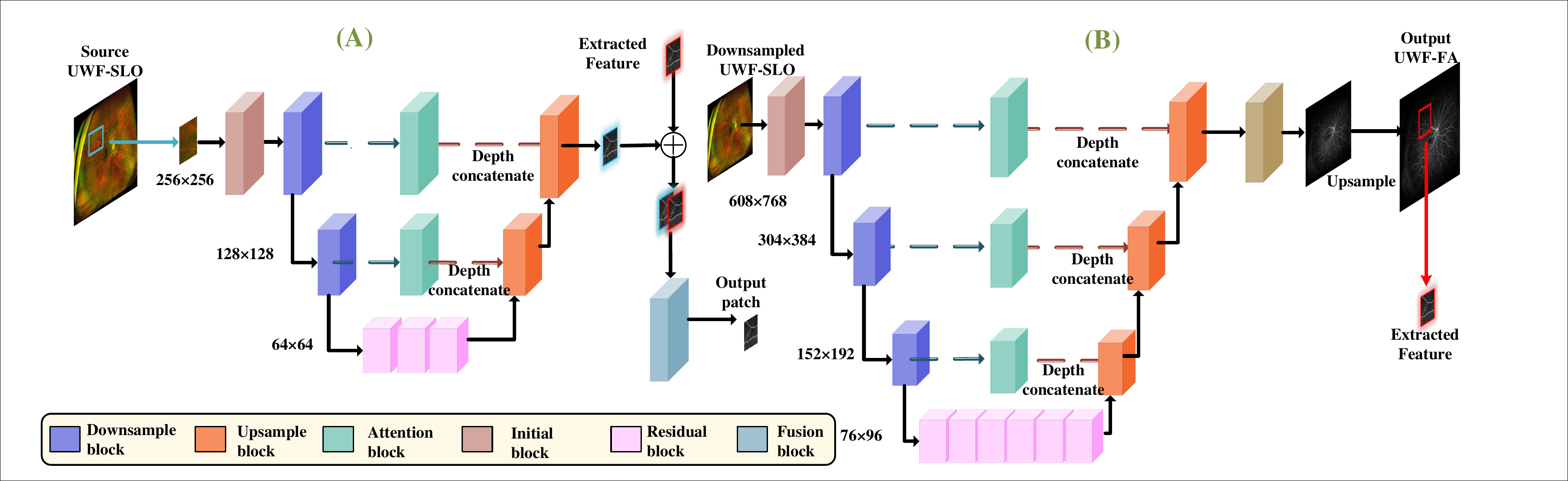}
\caption{Details of two-level generators and their building blocks. (A) $Gen_F$ consists of two down and up processes, and ultimately sends its resulting patch to $Gen_C$; (B) $Gen_C$  comprises of three down and up processes before sending its resulting patch to $Gen_F$.}
\label{2}
\end{figure}

\subsection{Patch and Fusion Module}\label{S2}
In Fig. \ref{1}, $Gen_F$ receives a randomly cropped patch as the input instead of the original image. This is because directly inputting the original image would occupy a large amount of memory and significantly reduce the training speed. Therefore, we only feed one cropped patch of the original image to $Gen_F$ in each step. In Fig. \ref{2}(A), a fusion block takes both patches from $Gen_F$ and $Gen_C$. To get the same region from $Gen_F$ and $Gen_C$, we resized the images to the same size, and cropped the patches from the same position. These two patches are concatenated at the depth level and passed into a two-layer fusion operation. 
\subsection{Attention Transmit Module}\label{S3}
In Fig. \ref{2}, the attention module is designed based on U-Net-like\cite{ref_16} architecture, whose sampling process can provide more information to the decoder. Whereas, when 
\begin{figure}
\centering 
\includegraphics[height=5.2cm,width=12cm]{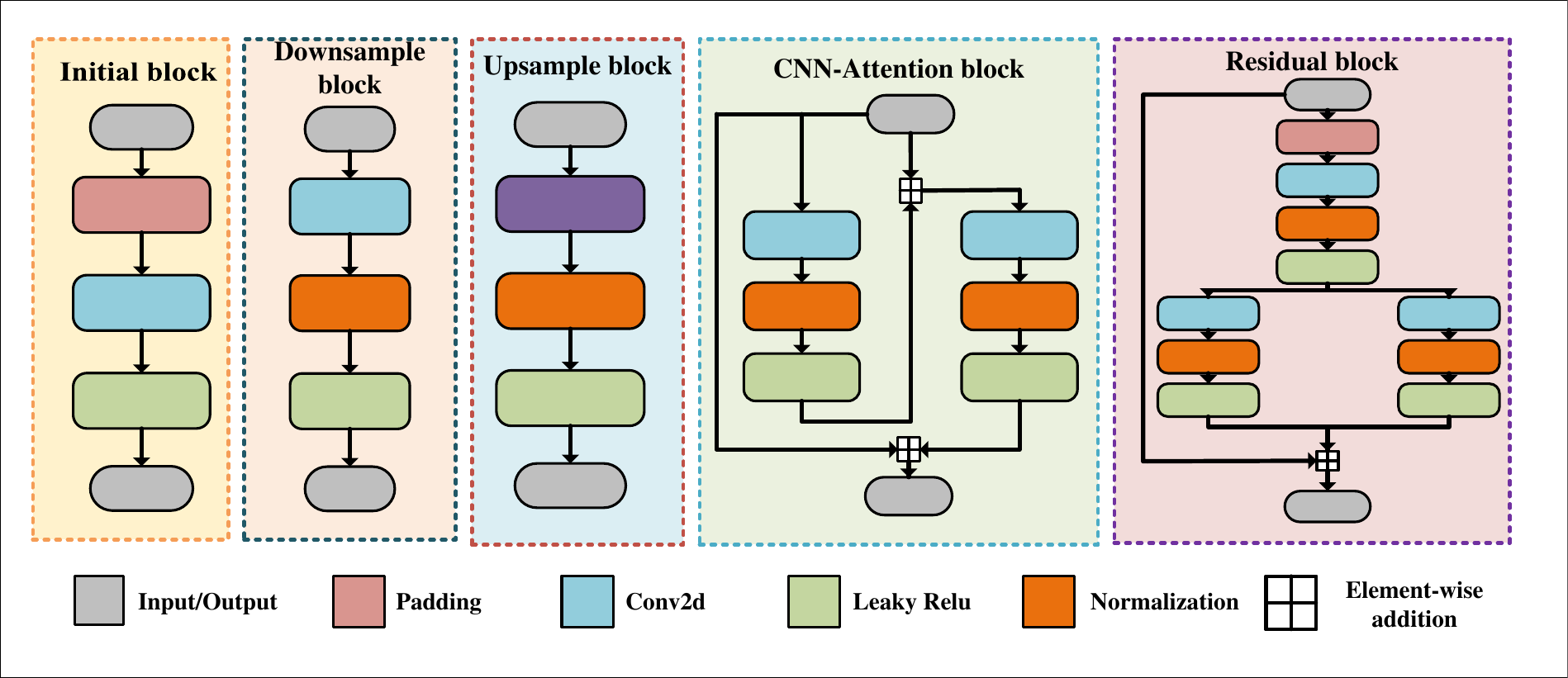}
\caption{Different blocks of the proposed generator. From left to right; the initial block, downsampling block, upsampling block, CNN-Attention block, and residual block.}
\label{3}
\end{figure}
synthesizing UWF-FA from UWF-SLO, the information density of the source image is low. For instance, the eye sockets in the periphery of the image are sparsely distributed blood vessels in some areas. Therefore, completely passing the graphs from the down-sampling process to the up-sampling process is not appropriate. Subsequently, images that pass through Attention Transmit Module can first extract useful information so that the decoder uses this information, efficiently. The multi-head attention \cite{ref_19} and the CNN-Attention blocks are shown in Fig. \ref{3}.

\subsection{Generator and Discriminator Architectures}\label{S4}
After conducting multiple experiments, we choose three down-sample layers for $Gen_F$ and two down-sample layers for $Gen_C$. In addition, each generator includes an initial block, down-sample block, up-sample block, residual block, and attention transmit module, which are shown in Fig. \ref{3}. The initial block contains a reflection padding, a 2D  convolution layer, and the Leaky-Relu activation function. The down/up-sampling blocks consist of a 2D convolution/transposed layer and the activation function combined with normalization. Additionally, the multi-scale discriminator in pix2pixHD \cite{ref_20} is employed to evaluate the output of the generator. For generator $Gen_X$, the first discriminator $D_{X1}$ takes the original and generated image $P_1$ while the second discriminator $D_{X2}$ takes the down-sampled version of $P_1$. Although theoretically, a multi-level discriminator can be applied by generating an image pyramid for an image, we use $D_{C1}$ and $D_{C2}$ for $Gen_C$, and $D_{F}$ for $Gen_F$ in our framework.

\subsection{Proposed Loss Function}\label{S5}
 Denote the two generators $Gen_C$ and $Gen_F$ as $G_C$ and $G_F$, the three discriminators as $D_{C1}$, $D_{C2}$, and $D_{F}$, and the paired variables \{($c_i, x_i$)\}, where $c$ represents the distribution of original input as a condition and $x$ represents the distribution of ground truth (i.e., real UWF-FA image). Given the conditional distribution $c$, we aim to maximize the loss of $D_{C1}$, $D_{C2}$, and $D_{F}$ while minimizing the loss of $Gen_C$ and $Gen_F$ using the following objective function:
\begin{equation}\label{B}
    \min_{G_C}\max_{D_{C1},D_{C2}}\sum_{k=1,2}\mathcal{L}_{cGAN}(G_C, D_{Ck}) + \min_{Gen_F}\max_{D_{F}}\mathcal{L}_{cGAN}(G_F, D_{F}),
\end{equation}
where $\mathcal{L}_{cGAN}$ is given by:
\begin{equation}
    \mathbb{E}_{(c,x)}[log(D(c,x))]+\mathbb{E}_c[log(1-D(G(c),c)].
\end{equation}
We adopt the feature mapping (FM) loss\cite{ref_20} in our framework. Firstly, we collect the target images and their translated counterparts as a pair of images. Then, we split the discriminators into multiple layers and obtain the output from each layer. Denote $D^{(i)}$ as the $ith$-layer to extract the feature, the loss function is then defined as:
\begin{equation}\label{C}
    \mathcal{L}_{FM}(G.D_k) = \mathbb{E}_{(c,x)}\sum_{i=1}^T\frac{1}{N_i}[\left\|D^i_k(c,x)-D^i_k(c,G(c))\right\|_1],
\end{equation}
where $T$ is the total number of layers and $N_i$ represents each layer's  number  of elements. (e.g., convolution, normalization, Leaky-Relu means three elements). Minimizing this loss ensures that each layer can extract the same features from the paired images. Additionally, we use the perceptual loss \cite{ref_8} in our framework, which is utilized by a pretrained VGG19 network \cite{ref_17}, to extract the features from the paired images and it is defined as:
\begin{equation}\label{D}
    \mathcal{L}_{VGG}(G.D_k) = \sum^N_{i=1}\frac{1}{M_i}[\left\|V^i(c,x)-V^i(c,G(c))\right\|_1],
\end{equation}
where $N$ represents the total number of layers, $M_i$ denotes the elements in each layer, and $V^i$ is the $ith$-layer of the VGG19 network. The final cost function is as follows:

\begin{equation}\label{E}
    \begin{split}
        \min_{G_C}(\max_{D_{C1},D_{C2}}\sum_{k=1,2}\mathcal{L}_{cGAN}(G_C, D_{Ck}) +   \lambda_{FMC}\sum_{k=1,2}\mathcal{L}_{FM}(G_C, D_{Ck})  +  \\ \lambda_{VGGC}\sum_{k=1,2}\mathcal{L}_{VGG}(G_C, D_{Ck})) 
    +  \min_{G_F}(\max_{D_F}\mathcal{L}_{cGAN}(G_F, D_{F}) + \\ \lambda_{FMF}\mathcal{L}_{FM}(G_F, D_{F}) +  \lambda_{VGGF}\mathcal{L}_{VGG}(G_F, D_{F})).
    \end{split}
\end{equation}
where $\lambda_{FMC},\lambda_{VGGC},\lambda_{FMF},\lambda_{VGGF}$ indicate adjustable weight parameters.
\section{Experiments and Results}
\subsection{Data Preparation and Preprocess}\label{M1}
In our experiments, we utilized an in-house dataset of UWF images obtained from a local hospital, comprising UWF-FA and UWF-SLO images. The UWF-SLO are in 3-channel RGB format, whereas the UWF-FA images are in 1-channel format. Each image pair was collected from a unique patient. However, from a clinical perspective, images taken with an interval of more than one day or those with noticeable fresh bleeding were excluded. Additionally, images that contain numerous interfering factors affecting their quality were also discarded. After the quality check, we have 70 paired images with the size of 3900×3072, of which 70\% were randomly allocated for training and 30\% for testing, respectively. \par
Furthermore, we employed image sharpening through histogram equalization to enhance the clarity of images. We then utilized automated image registration software, i2k Retina Pro, to register each pair of images which changed the image size. To standardize the size of each image, we resized the registered images to 2432×3702. Subsequently, we randomly cropped the resized images with a size of 608×768 into different patches. And 50 patches could be obtained for each image. Finally, we adopted data augmentation using random flip and rotation to increase the number of training images from 49 pairs to 1960 pairs. 

\subsection{Implementation Details}\label{M2}
All our experiments were conducted on the PyTorch 1.12 framework and carried out on two Nvidia RTX 3090Ti GPUs. Our model was trained from scratch to 200 epochs. The parameters were optimized by the Adam optimizer algorithm \cite{ref_13} with learning rate $\alpha$ = 0.0002, $\beta_1$ = 0.5 and  $\beta_2$ = 0.999. We used a batch size of 2 to train our model and set  $\lambda_{FMF}=\lambda_{FMC}=\lambda_{VGGF}=\lambda_{VGGC}$= 10 (Eq. \ref{E}).
\subsection{Comparisons}\label{M3}
We first compared the performance of our model with some state-of-the-art GAN-based models including: Pix2pix \cite{ref_7}, Pix2pixHD \cite{ref_20} and StarGAN-v2 \cite{ref_3}. For a fair comparison, we took the default parameters of the open-source codes of the competing methods, ensuring that the data volume matched the number of training cycles. We used the $FID(\downarrow), KID(\downarrow), LPIPS(\downarrow)$ and $IS(\uparrow)$ to evaluate the generated UWF-FA. Table \ref{T1} shows the generation performance of different methods. Overall, our method achieves the best in all metrics compared to other models. The Pix2Pix attained the worst performance in all evaluation metrics while PixPixHD and StarGAN had comparable performance. In general, our method outperformed the competing methods and improved FID, KID, IS, and LPIPS by at least 24.47\%, 39.95\%, 3.59\%, and 14.04\%, respectively. Although StarGAN-v2 yielded the second-best performance, it is still less comparable with the proposed UWAT-GAN due to the lesion generation module which could capture tiny image details and improve overall performance (see Fig. \ref{4}).

\begin{table}[t]
\caption{Comparison with the state-of-the-art methods using 4 evaluation metrics. The * means that the official code hasn't provided the way to measure it. }
\centering  
\begin{tabular}{|p{3cm}<{\centering}|p{2cm}<{\centering}|p{2cm}<{\centering}|p{2cm}<{\centering}|p{2cm}<{\centering}|}
\hline
Methods & FID($\downarrow$) & KID($\downarrow$) & IS($\uparrow$) & LPIPS($\downarrow$) \\
\hline
Pix2Pix \cite{ref_7} & 135.4038 & 0.1094 & 1.2772 & 0.4575 \\
\hline
Pix2PixHD \cite{ref_20} & 76.76 & 0.0491 & 1.0602 & 0.4451\\
\hline
StarGAN-v2 \cite{ref_3}& 74.38 & 0.0433 & * & 0.4577\\
\hline
UWAT-GAN $M_{NA}$ & 67.96 & 0.0308 & 1.2757 & 0.4086 \\
\hline
\textbf{UWAT-GAN} & \textbf{55.59} & \textbf{0.0260} & \textbf{1.323} & \textbf{0.3826}\\
\hline 
\end{tabular}
\label{T1}
\end{table}
\subsection{Ablation Study}\label{M3}
\begin{figure}[t]
\centering
\includegraphics[width=10cm, height = 4cm]{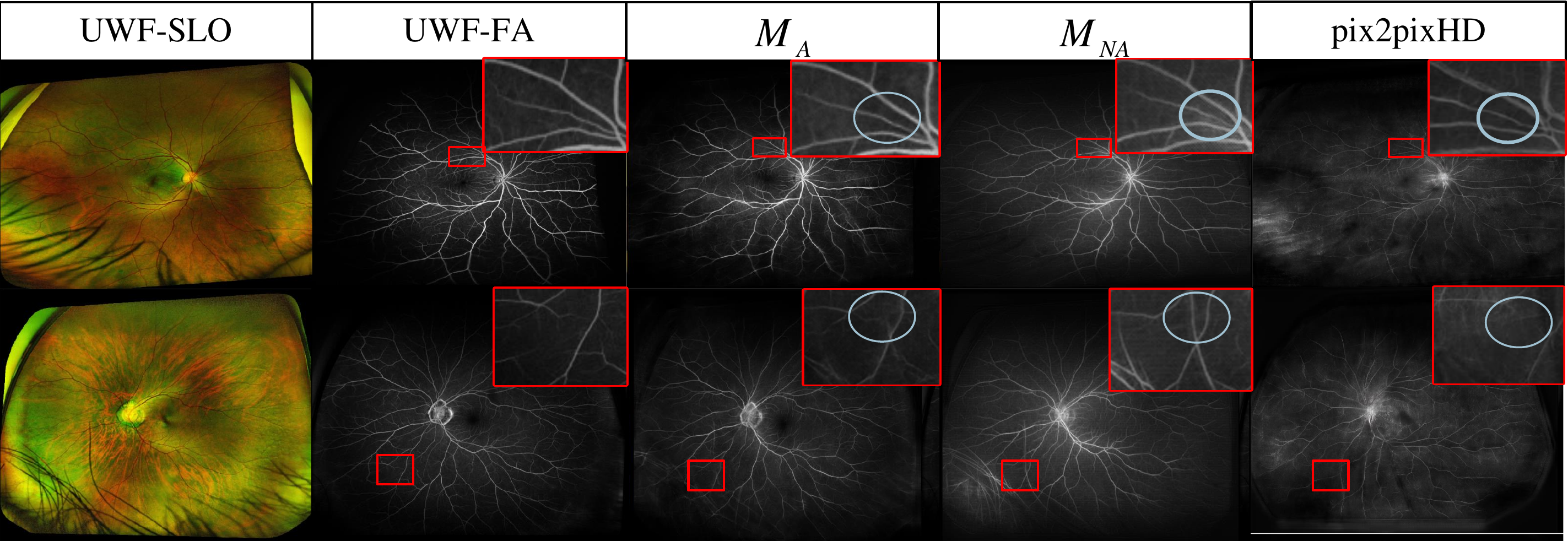}
\caption{Visualization of original and generated images. From left to right: source UWF-SLO, UWF-FA, the proposed framework with and without the attention transmit module, and the pix2pixHD, respectively. }
\label{4}
\end{figure}

To evaluate the significance of the attention transmit module proposed in UWAT-GAN, we trained our model with and without this module, namely $M_A$ and $M_{NA}$, respectively. Unlike the generated images of $M_A$, we found that $M_{NA}$ was not so distinctive as some vessels were missing and some interference of eyelashes was incorrectly considered as vessels. In Fig. \ref{4}, we showed the original pair of UWF-SLO and UWF-FA images and the generated images with $M_A$ and $M_{NA}$. It is clear that the proposed method can generate good images and preserve small details. It becomes more distinctive when the attention module was used, as shown in the enlarged view of the red rectangle. It is also obvious that the FID and KID scores were improved by 22.25\% and 18.46\%, respectively.

\section{Discussion and Conclusion}
To address the potential adverse effects of fluorescein injection during FA, we propose UWAT-GAN to synthesize UWF-FA from UWF-SLO. Our method can generate high-resolution images and enhance the ability to capture small vascular lesions. Comparison and ablation study on an in-house dataset demonstrate the superiority and effectiveness of our proposed method. However, our model still has a few limitations. First, not every pair of images can be registered since some paired images may have fewer available features, making registration difficult. Second, our model's accuracy in synthesizing very tiny lesions is not optimal, as some lesions cannot be well generalized. Third, the limited size of our dataset is relatively small and may affect the model performance. In the future, we aim to expand the size of our dataset and explore the use of the object detection model, especially for small targets, to push our model pay more attention to some lesions. After further validation, we aim to adopt this method as an auxiliary tool to diagnose and detect fundus diseases.

\subsubsection{Data statement} Dataset used in this work is privately collected from our collaborative hospital after ethical approval and data usage permission. Data maybe be available upon the permission of the related authority and adequate request to the corresponding author(s).
\subsubsection{Acknowledgements} This work was supported by the National Natural Science Foundation of China (No.U20A20386, U22A2033), Zhejiang Provincial Natural Science Foundation of China (No.LY21F020017),  Chinese Key-Area Research and Development Program of Guangdong Province (2020B0101350001), GuangDong Basic and Applied Basic Research Foundation (No. 2022A1515110570), Innovation teams of youth innovation in science and technology of high education institutions of Shandong province (No. 2021KJ088), the Shenzhen Science and Technology Program (JCYJ20220818103001002), and the Guangdong Provincial Key Laboratory of Big Data Computing, The Chinese University of Hong Kong, Shenzhen. 

\bibliographystyle{splncs04}
\bibliography{paper2942} 

\begin{thebibliography}{10}
\providecommand{\url}[1]{\texttt{#1}}
\providecommand{\urlprefix}{URL }
\providecommand{\doi}[1]{https://doi.org/#1}

\bibitem{ref_1}
Ashraf, M., Shokrollahi, S., Salongcay, R.P., Aiello, L.P., Silva, P.S.:
  Diabetic retinopathy and ultrawide field imaging. In: Seminars in
  Ophthalmology. vol.~35, pp. 56--65. Taylor \& Francis (2020)

\bibitem{ref_2}
Barratt, S., Sharma, R.: A note on the inception score. arXiv preprint
  arXiv:1801.01973  (2018), \url{http://arxiv.org/abs/1801.01973}

\bibitem{ref_3}
Choi, Y., Uh, Y., Yoo, J., Ha, J.W.: Stargan v2: Diverse image synthesis for
  multiple domains. In: Proceedings of the IEEE/CVF conference on computer
  vision and pattern recognition. pp. 8188--8197 (2020)

\bibitem{ref_4}
Ehlers, J.P., Jiang, A.C., Boss, J.D., Hu, M., Figueiredo, N., Babiuch, A.,
  Talcott, K., Sharma, S., Hach, J., Le, T., et~al.: Quantitative
  ultra-widefield angiography and diabetic retinopathy severity: an assessment
  of panretinal leakage index, ischemic index and microaneurysm count.
  Ophthalmology  \textbf{126}(11),  1527--1532 (2019)

\bibitem{ref_5}
Goodfellow, I., Pouget-Abadie, J., Mirza, M., Xu, B., Warde-Farley, D., Ozair,
  S., Courville, A., Bengio, Y.: Generative adversarial networks.
  Communications of the ACM  \textbf{63}(11),  139--144 (2020)

\bibitem{ref_6}
Heusel, M., Ramsauer, H., Unterthiner, T., Nessler, B., Hochreiter, S.: Gans
  trained by a two time-scale update rule converge to a local nash equilibrium.
  Advances in neural information processing systems  \textbf{30} (2017)

\bibitem{ref_7}
Isola, P., Zhu, J.Y., Zhou, T., Efros, A.A.: Image-to-image translation with
  conditional adversarial networks. In: Proceedings of the IEEE conference on
  computer vision and pattern recognition. pp. 1125--1134 (2017)

\bibitem{ref_8}
Johnson, J., Alahi, A., Fei-Fei, L.: Perceptual losses for real-time style
  transfer and super-resolution. In: Computer Vision--ECCV 2016: 14th European
  Conference, Amsterdam, The Netherlands, October 11-14, 2016, Proceedings,
  Part II 14. pp. 694--711. Springer (2016)

\bibitem{ref_9}
Kamran, S.A., Hossain, K.F., Tavakkoli, A., Zuckerbrod, S.L., Baker, S.A.:
  Vtgan: Semi-supervised retinal image synthesis and disease prediction using
  vision transformers. In: Proceedings of the IEEE/CVF international conference
  on computer vision. pp. 3235--3245 (2021)

\bibitem{ref_10}
Kamran, S.A., Hossain, K.F., Tavakkoli, A., Zuckerbrod, S.L., Sanders, K.M.,
  Baker, S.A.: Rv-gan: Segmenting retinal vascular structure in fundus
  photographs using a novel multi-scale generative adversarial network. In:
  Medical Image Computing and Computer Assisted Intervention--MICCAI 2021: 24th
  International Conference, Strasbourg, France, September 27--October 1, 2021,
  Proceedings, Part VIII 24. pp. 34--44. Springer (2021)

\bibitem{ref_11}
Knop, S., Mazur, M., Spurek, P., Tabor, J., Podolak, I.: Generative models with
  kernel distance in data space. Neurocomputing  \textbf{487},  119--129 (2022)

\bibitem{ref_12}
Li, C., Wand, M.: Precomputed real-time texture synthesis with markovian
  generative adversarial networks. In: Computer Vision--ECCV 2016: 14th
  European Conference, Amsterdam, The Netherlands, October 11-14, 2016,
  Proceedings, Part III 14. pp. 702--716. Springer (2016)

\bibitem{ref_13}
Lihua, L.: Simulation physics-informed deep neural network by adaptive adam
  optimization method to perform a comparative study of the system. Engineering
  with Computers  \textbf{38}(Suppl 2),  1111--1130 (2022)

\bibitem{ref_14}
Liu, X., Yu, A., Wei, X., Pan, Z., Tang, J.: Multimodal mr image synthesis
  using gradient prior and adversarial learning. IEEE Journal of Selected
  Topics in Signal Processing  \textbf{14}(6),  1176--1188 (2020)

\bibitem{ref_15}
Luo, S.: A survey on multimodal deep learning for image synthesis:
  Applications, methods, datasets, evaluation metrics, and results comparison.
  In: 2021 the 5th International Conference on Innovation in Artificial
  Intelligence. pp. 108--120 (2021)

\bibitem{ref_16}
Ronneberger, O., Fischer, P., Brox, T.: U-net: Convolutional networks for
  biomedical image segmentation. In: Medical Image Computing and
  Computer-Assisted Intervention--MICCAI 2015: 18th International Conference,
  Munich, Germany, October 5-9, 2015, Proceedings, Part III 18. pp. 234--241.
  Springer (2015)

\bibitem{ref_17}
Simonyan, K., Zisserman, A.: Very deep convolutional networks for large-scale
  image recognition. arXiv preprint arXiv:1409.1556  (2014),
  \url{http://arxiv.org/abs/1409.1556}

\bibitem{ref_18}
Tavakkoli, A., Kamran, S.A., Hossain, K.F., Zuckerbrod, S.L.: A novel deep
  learning conditional generative adversarial network for producing angiography
  images from retinal fundus photographs. Scientific Reports  \textbf{10}(1),
  21580 (2020)

\bibitem{ref_19}
Vaswani, A., Shazeer, N., Parmar, N., Uszkoreit, J., Jones, L., Gomez, A.N.,
  Kaiser, {\L}., Polosukhin, I.: Attention is all you need. Advances in neural
  information processing systems  \textbf{30} (2017)

\bibitem{ref_20}
Wang, T.C., Liu, M.Y., Zhu, J.Y., Tao, A., Kautz, J., Catanzaro, B.:
  High-resolution image synthesis and semantic manipulation with conditional
  gans. In: Proceedings of the IEEE conference on computer vision and pattern
  recognition. pp. 8798--8807 (2018)

\bibitem{ref_21}
Wang, X., Ji, Z., Ma, X., Zhang, Z., Yi, Z., Zheng, H., Fan, W., Chen, C.:
  Automated grading of diabetic retinopathy with ultra-widefield fluorescein
  angiography and deep learning. Journal of Diabetes Research  \textbf{2021}
  (2021)

\bibitem{ref_22}
Xiao, Y., Peters, K.R., Fox, W.C., Rees, J.H., Rajderkar, D.A., Arreola, M.M.,
  Barreto, I., Bolch, W.E., Fang, R.: Transfer-gan: multimodal ct image
  super-resolution via transfer generative adversarial networks. In: 2020 IEEE
  17th International Symposium on Biomedical Imaging (ISBI). pp. 195--198. IEEE
  (2020)

\bibitem{ref_23}
Yang, Q., Li, N., Zhao, Z., Fan, X., Chang, E.I.C., Xu, Y.: Mri cross-modality
  image-to-image translation. Scientific reports  \textbf{10}(1), ~3753 (2020)

\bibitem{ref_24}
Zhang, R., Isola, P., Efros, A.A., Shechtman, E., Wang, O.: The unreasonable
  effectiveness of deep features as a perceptual metric. In: Proceedings of the
  IEEE conference on computer vision and pattern recognition. pp. 586--595
  (2018)

\end{thebibliography}

\end{document}